\documentstyle[aaspp4,12pt]{article}
\begin{document}

\def\P{\bar{\Phi}}
\def\st{\sigma_{\rm T}}
\def\vk{v_{\rm K}}
\def\sles{\lower2pt\hbox{$\buildrel {\scriptstyle <}
   \over {\scriptstyle\sim}$}}
\def\sgreat{\lower2pt\hbox{$\buildrel {\scriptstyle >}
   \over {\scriptstyle\sim}$}}

\title{A Cosmic Battery}

\author{Ioannis Contopoulos \altaffilmark{1} and
Demosthenes Kazanas \altaffilmark{2}}
\affil{}
\altaffiltext{1}{Physics Department, University of
Crete, P. O. Box 2208, Heraklion 71003, Greece}
\altaffiltext{2}{NASA/Goddard Space Flight Center, Greenbelt, MD 20771}

\begin{abstract}

We show that the Poynting--Robertson drag effect in an optically thin
advection--dominated accretion flow around active gravitating objects
generates strong azimuthal electric currents which give rise to
astrophysically significant magnetic fields. Although the mechanism is
most effective in accreting compact objects, it seems very promising
to also account for the generation of stellar dipolar fields during
the late protostellar collapse phase, when the star approaches the
main sequence.

\end{abstract}

\section{Introduction}

The origin of cosmic magnetic fields remains one of the open questions
in astrophysics.  Their origin has been sought in the entire history
of the Universe going as far back as the QCD phase transition (Sigl,
Olinto \& Jedamzik~1997) the electroweak phase transition (e.g. Baym,
Bodeker \& McLerran~1996; Joyce \& Shaposhnikov~1997) or even close to
the Planck time (e.g. Ratra 1992).  Most theories attempt to produce
the observed magnetic fields by starting with some `seed magnetic
field' and follow its amplification by means of a `dynamo mechanism',
i.e.  a mechanism which can convert kinetic energy of the conducting
fluid, into which the magnetic field is supposedly frozen, into
magnetic field energy density (Moffat~1978; Parker~1979).
Unfortunately, when one considers the back reaction of the stretched
and folded magnetic field on the dynamics of the conducting fluid,
dynamo action seems to be very ineffective, in that as the field at
the small diffusive scales reaches equipartition, the large-scale
component remains several orders of magnitude weaker than
astrophysical observed magnetic fields (Vainshtein \& Cattaneo~1992).

In addition to these problems with the effectiveness of the dynamo
theory, the origin of the `seed magnetic field' needed to be amplified
is also an issue, assumed in most cases as a boundary condition.
However, the consideration that in its initial state the Universe was
homogeneous and isotropic to a high degree, excludes the presence of
large scale magnetic fields. The seed magnetic field is therefore
considered to be generated by some `battery mechanism', the most
common being the `Biermann battery' (Biermann~1950), i.e. a
mechanism for generating large scale electric currents based
%on the misalignment of the pressure and density gradients in a plasma
on the thermoelectric effect.  In this respect it is worth reminding
the reader that any battery mechanism depends on the possibility of
imparting and sustaining different velocities to the electrons and
protons in a plasma, thus creating an electric current. Obviously, in
order for such an electric current to give rise to a steady (rather
than transient) magnetic field it is necessary that the line integral
of the proton-electron velocity difference along a closed, large scale
circuit does not vanish.  This leads directly to the notion that the
generation of magnetic fields should be naturally associated with
motions in which the integral $\oint ${\bf v}$ \cdot d${\bf l}
%$\oint v\cdot ds$ 
does not vanish, i.e. motions with
non-zero circulation, a quantity also thought to be zero at the early
Universe but generated later, in situations which involve the presence
of dissipation, thus suggesting dissipation as a necessary condition for
the generation of large scale magnetic fields too.

Herein we revisit the theory of battery generated cosmic magnetic
fields by employing a specific mechanism for producing such currents
in a specific astrophysical setting, namely the Poynting--Robertson
drag effect on the electrons of an Advection--Dominated Accretion Flow
(Narayan \& Yi~1994, 95; hereafter ADAF) in the vicinity of an
accreting object. We examine the time evolution of these magnetic
fields in a variety of several astrophysical settings.  We find that
this effect may be significant enough, at some phase in the evolution
of the accreting object, as to not require any further amplification
in order to account for the observed, present day values associated
with the magnetic fields of the class of objects in question.

In \S~2 we give a brief outline of our mechanism and examine
qualitatively its implications and effects in different astrophysical
settings. In \S~3 we integrate numerically the associated equations and
compare our results to those derived earlier on qualitative basis.
Finally, in \S~4 the results are summarized and conclusions are drawn.

\section{The model}

The model is simple: Consider a point-like astrophysical energy source
of luminosity $L$, and an optically thin, fully ionized ADAF with
rotational velocity $v_\phi=Ar\Omega_{\rm K}$ at distances $r_{\rm
in}\leq r \leq r_{\rm out}$ around the above energy source
($\Omega_{\rm K}$ is the Keplerian angular velocity, and $A<1$ is a
constant parameter of the flow).  Assuming that its scattering depth
is sufficiently small that it scatters photons from the center only
once, each plasma electron feels an azimuthal force equal to
\begin{equation}
F_{\rm P-R} = -\frac{L\sigma_{\rm T}}{4\pi r^2 c}\frac{v_\phi}{c}
\label{FPR}
\end{equation}
against its direction of rotation in the region between $r_{\rm in}$
and $r_{\rm out}$ where the plasma is optically thin.  This is the
Poynting--Robertson radiation drag.  The positive ions feel a force
$(m_i/m_e)^2$ times smaller which can be neglected. As a consequence
of the above force, the electrons tend to lag behind the protons, and
an azimuthal electric current density 
\begin{equation}
J=\frac{c}{4\pi}\left(\frac{\partial B_r}{\partial z}-
\frac{\partial B_z}{\partial r}\right) \sim
-\frac{c}{4\pi}\frac{\partial B_z}{\partial r}
\label{J}
\end{equation}
will develop\footnote{In order to keep the analysis simple, we will
assume a cylindrical distribution for the magnetic field inside the
geometrically thick ADAF, i.e.  we will neglect its radial component
and the associated magnetic field tension.}\footnote{The reader should
not worry that `flux--freezing' is not satisfied as electrons and ions
move at slightly different azimuthal velocities.  Electric currents
generate magnetic forces which balance the forces ${\bf F}_e$ and
${\bf F}_i$ acting on electrons and ions respectively. The reader can
convince him/herself that when the problem is solved, the two plasma
constituents perform ${\bf F}_{e,i}\times {\bf B}/(q_{e,i}B^2)$ drifts
across the above force fields, i.e. they do not move together even
under conditions of flux--freezing (this analysis remains outside the
main scope of the present work, and can be found in Ciolek \&
Mouschovias~1993).}. Now, during the ADAF phase, the plasma is
turbulent with a turbulent plasma conductivity $\sigma_{t}$ (see
below), whereas when the ADAF phase ceases, the plasma conductivity
attains a much higher value, $\sigma_{\rm Spitzer}=10^{17}\ T_6^{1.5}\
{\rm s}^{-1}$ (e.g. Zombeck~1992; $T_n$ is the plasma temperature in
units of $10^n$~K). On average, the effect of the finite plasma
conductivity $\sigma_{\rm eff}$ can be parametrized as an extra
azimuthal drag force equal to
\begin{equation}
F_{\rm drag} = \frac{Je}{\sigma_{\rm eff}}\ ,
\label{Fdrag}
\end{equation}
acting on the electrons (similarly for the ions).  The azimuthal
component of the microscopic momentum (or force) equation for the
electrons can be written as
\begin{equation}
-e(E - \frac{v_r}{c}B_z) + F_{\rm P-R} + F_{\rm drag}=0\ ,
\label{E}
\end{equation}
where $E$ is the azimuthal component of the electric field; and $v_r$
is the radial flow velocity; $e$ is taken positive.  Combining
eqs.~(\ref{FPR}), (\ref{J}), (\ref{Fdrag}), and (\ref{E}) the axial
component of the induction equation $\partial B_z/\partial t =
-(c/r)\partial (rE)/\partial r$ yields
\begin{equation}
\frac{\partial B_z}{\partial t} = 
\frac{1}{r}\frac{\partial}{\partial r}\left\{
\frac{L\st A \Omega_{\rm K}}{4\pi ce}
-r v_r B_z + \frac{c^2}{4 \pi \sigma_{\rm eff}}r
\frac{\partial B_z}{\partial r}
\right\}\ ,
\label{induction}
\end{equation}
in the range of radii $(r_{\rm in},r_{\rm out})$.  $\Psi$ is defined
through the relation $B_z=r^{-1}\partial\Psi/\partial r$, and
$2\pi\dot{\Psi}$ is the growth rate of the total magnetic flux
contained within radius $r$.  One should note that
eq.~(\ref{induction}) includes both the advection of magnetic field
through the term $-r v_r B_z$, as well as its diffusion through the
term $c^2/4 \pi \sigma_{\rm eff} (r\partial B/
\partial r)$. The connection between magnetic field and circulation 
discussed in the introduction, then, suggests that in an accretion
disk, the turbulent magnetic diffusivity should also be proportional to
the quantity effecting the dissipation in the accretion disk, namely
the turbulent viscosity parameter. Thus one may write
(e.g. Reyes--Ruiz \& Stepinski~1996)
\begin{equation}
\eta_{\rm eff} \equiv 
c^2/(4\pi\sigma_{\rm eff})
\sim {\cal P}_m\nu 
\sim {\cal P}_m\alpha \xi r^2\Omega_{\rm K}\ ,
\label{eta}
\end{equation}
where, $\alpha<1$ (the usual Shakura--Sunyaev (1973) $\alpha$
parameter) and $\xi<1$ (the ratio of the local flow temperature to the
virial temperature) are taken to be constant flow parameters (we
remind the reader that $A\equiv\Omega/\Omega_{\rm K}=(1-5\xi/2
-\alpha^2\xi^2/2)^{1/2}$; Narayan \& Yi~1994, 95); ${\cal P}_m$ is the
ratio of magnetic diffusivity to viscosity, the so-called
magnetic Prandtl number.  Note in addition that in an ADAF, the 
radial velocity can be related to the azimuthal one through 
the relation
\begin{equation}
v_r \sim -\alpha\xi r\Omega_K \sim -\eta_{\rm eff}/({\cal P}_m r)\ .
\label{vr}
\end{equation}
Substituting all the above in the induction equation leads to 
\begin{equation}
\frac{\partial B_z}{\partial t} = 
\frac{1}{r}\frac{\partial}{\partial r}\left\{
\frac{L\st A \Omega_{\rm K}}{4\pi ce} 
+ r^2 \Omega_{\rm K} 
\alpha \xi \left[B_z+{\cal P}_m r\frac{\partial B_z}{\partial r}\right]
\right\}\equiv
\frac{1}{r}\frac{\partial \dot{\Psi}}{\partial r}
\label{inductionB}
\end{equation}

The r.h.s. of eq.~(\ref{inductionB}) consists of a source term
(independent of $\sigma_{\rm eff}$), an advection term, and a
diffusion term. The source term is the Poynting-Robertson azimuthal
electric current and leads to field growth which is fastest at
$r=r_{\rm in}$.  This gives rise to a distribution of closed poloidal
magnetic field loops, with their center somewhere inside $r_{\rm in}$.
As we will see in the next subsection, {\em when ${\cal P}_m\sgreat
2$, the magnetic flux contained inside $r_{\rm in}$ grows almost
linearly in time, as long as the ADAF persists. When rotation and
accretion cease, and $\sigma_{\rm eff}\sim \sigma_{\rm Spitzer}$, this
flux will remain effectively trapped in the plasma over timescales}
\begin{equation}
\tau \simeq \frac{r_{\rm in}^2}{\eta_{\rm Spitzer}} \simeq
4 x^2 M_0^2 T_6^{1.5}\ {\rm yr}\ .
\label{time}
\end{equation}
$M_n$ is the mass of the central object in units of $10^n~M_\odot$;
$x\equiv r/r_{\rm S}$ is the radial coordinate in units of the central
object's Schwarzschild radius $r_{\rm S}$.

Obviously, this mechanism will cease being effective when the magnetic
field becomes strong enough to affect the dynamics of accretion.  In
addition, a magnetic field which becomes dynamically important will
also affect the torques from the ADAF onto the star (or vice
versa). We prefer to defer discussion of the implications of this
aspect of the problem and the rotational evolution of the compact
object to a future publication; as indicated by the present analysis
these could indeed be significant.

It is of interest to examine the time scales necessary to accumulate
enough flux via the above mechanism in the most optimistic case with
${\cal P}_m>2$, for the corresponding magnetic field to affect
significantly the dynamics of accretion, as this may be an important
component to the formation of jets, which are ubiquitous in accreting
compact objects.  Assuming an almost constant growth rate of magnetic
flux $\Psi$ of the order of the first term in the r.h.s. of
eq.~(\ref{induction}),
\begin{equation}
\Psi \sim 
3\cdot 10^{23} \, \frac{f}{x^{3/2}}  
\left(\frac{t}{{\rm yrs}}\right) ~~~ {\rm G ~cm}^2\ ,
\label{psixt}
\end{equation}
%\begin{eqnarray}
%\Psi & \sim & 
%%1.5 \times 
%10^{12} \, M_0 \, f \;
%\frac{t}{t_{\rm ff}} 
%\sim 
%%1.5 \times 
%10^{17} \, \frac{f}{x^{3/2}} 
%\left(\frac{t}{{\rm sec}}\right) ~~~ {\rm G ~cm}^2 \nonumber\\
%& \sim & 
%%4.7 \times 
%5\cdot 10^{24} \, \frac{f}{x^{3/2}}  
%\left(\frac{t}{{\rm yrs}}\right) ~~~ {\rm G ~cm}^2\ ,
%\label{psixt}
%\end{eqnarray}
where, $f$ is the fraction of the source luminosity compared to the
Eddington luminosity.  One should note that the latter expression is
independent of the mass of the system and depends only on its
luminosity through the parameter $f$ (for compact objects $x$ is
considered to be of order unity). However, the magnetic field itself
is bounded by considerations associated with equipartition of the
magnetic energy density with that of the ram pressure of almost
free-falling matter,
\begin{equation}
B_{eq} \sim
%8.8
9\cdot 10^7~x^{-1.25} {\dot m}^{0.5} M_0^{-0.5}~~{\rm G}\ ,
\label{Beq}
\end{equation}
where $\dot m$ is the accretion rate in units of the Eddington rate.
Folowing Narayan \& Yi~1994, we took $\alpha=0.3$, $\xi=0.38$, and
$A=0.2$.  Assuming further that $B(r\leq r_{\rm in})\sim
\Psi(r_{\rm in})/r_{\rm in}^2$, and that $f \simeq {\dot m}/x$, one
obtains
\begin{equation}
\tau_{eq} \sim  
%3.7
4\cdot 10^{-5} x^{3.25} M_0^{1.5} 
{\dot m}^{-0.5}~~{\rm yr}\ .
\end{equation}
This value is comparable to the Eddington time ($\sim 10^8$ yrs) for
an AGN with a massive central black hole ($M_9 \sim 1$), but decreases
very quickly with the decrease of the mass.

\section{Numerical Simulation}

The detailed time evolution of the magnetic field can be computed by
direct integration of the time evolution equation (\ref{induction}).
We introduce the dimensionless variables
\begin{equation}
\hat{r}=r/r_{\rm in}\ ,\ \hat{t}=t/t_o\ ,\ 
b=B/B_o\ ,\ \dot{\psi}=\dot{\Psi}t_o/(r_{\rm in}^2 B_o)\ ,
\end{equation}
where, $t_o\equiv (\Omega_{\rm K}(r_{\rm in})\alpha\xi)^{-1}$;$
B_o\equiv L\st A/(4\pi c e r_{\rm in}^2\alpha\xi)$. The induction
equation can now be rewritten in dimensionless form applicable in
various astrophysical situations,
\begin{equation}
\frac{\partial b}{\partial \hat{t}}=
\frac{1}{\hat{r}}\frac{\partial}{\partial\hat{r}}\left\{
\hat{r}^{-1.5}+\hat{r}^{0.5}\left[b+
{\cal P}_m \hat{r}\frac{\partial b}{\partial \hat{r}}\right]
\right\}\equiv
\frac{1}{\hat{r}}\frac{\partial \dot{\psi}}{\partial\hat{r}}\ .
\label{induction2}
\end{equation}

The induction equation is integrated numerically in the interval
$1\leq \hat{r} \leq \hat{r}_{\rm out}$. In order to keep the number of
free parameters to a minimum, we take $\hat{r}_{\rm out}\gg 1$.  We
follow the numerical procedure described in the Appendix of
Mouschovias \& Morton~(1991), and implement the `free boundary'
condition $\partial b/\partial\hat{r}=0$ at $\hat{r}=\hat{r}_{\rm
in}$, allowing thus any amount of magnetic flux that would be advected
inside $\hat{r}_{\rm in}$ to enter unimpeded.  We discovered that, for
magnetic Prandtl numbers ${\cal P}_m> 2$, the dimensionless magnetic
flux inside dimensionless radius $\hat{r}=1$ rises almost linearly
with time, with a dimensionless flux growth rate $\dot{\psi}$ of order
unity.  In fact, when ${\cal P}_m\geq 2$, eq.~(\ref{induction2}) has
the steady--state solution
\begin{equation}
b(\hat{r}) = \frac{1}{(2{\cal P}_m -1)}\left[
\frac{1}{\hat{r}^2}-\frac{4}{\hat{r}^{0.5}}\right]\ ,
\label{steadystate}
\end{equation}
with, $\dot{\psi} = 2({\cal P}_m-2)/(2{\cal P}_m -1)$.  We understand
this interesting behavior as follows: The Poynting--Robertson effect
generates an azimuthal electric current resulting in closed loops of
poloidal magnetic field with a distribution of positive magnetic flux
inside $r_{\rm in}$ (where the magnetic field points, say, upward) and
negative flux outside, i.e. in the regions with opposite field
polarity.  The rate of positive flux growth inside $r_{\rm in}$ is
given by the first term in the r.h.s. of eq.~(\ref{induction2}). When
${\cal P}_m\geq 2$, diffusion is dominant, and the distribution of
{\em negative} flux outside $r_{\rm in}$ diffuses outward to larger
distances. As a result, positive flux is generated without limit
(except of course for the equipartition limit) in the interior.  On
the other hand, when ${\cal P}_m<2$, the inward advection of the
negative flux dominates over its outward diffusion. As a result, the
steady--state positive flux accumulated inside $r_{\rm in}$ remains
finite, since its rate of growth by the Poynting--Robertson mechanism
is balanced by the rate of inward advection of the return (negative)
flux.

In figure~1 we exhibit the spatio-temporal evolution of the magnetic
field and the associated flux resulting from our numerical solution to
eq.~(\ref{induction2}). In figure~1a we show the evolution of the
radial distribution of the magnetic field as a function of the
dimensionless radius $r/r_{\rm in}$ for ${\cal P}_m = 2.5$. Each curve
corresponds to an increment in time by $10t_o$. In figure 1b we
exhibit the evolution of the magnetic flux held within $r_{\rm in}$ as
a function of time for different values of ${\cal P}_m$. The reader
should note that according to eq.~(\ref{steadystate}), for Prandtl
numbers ${\cal P}_m> 2$, a steady state situation, though it
corresponds to a constant value for the magnetic field in the region
$r/r_{\rm in}>1$, demands a linear increase for the magnetic flux
$\Psi$ for $r/r_{\rm in} < 1$, with the return flux diffusing steadily
to infinity.

In table~1, we introduce dimensions, and apply our mechanism to
specific astrophysical situations. The mechanism is most effective at
short scales, especially around compact objects (neutron stars, black
holes of both stellar and galactic sizes). One could also suggest that
magnetic flux generated quickly around active galactic nuclei, could
be carried by winds out to galactic (and even extragalactic) scales
(similar mechanisms have already been proposed in order to account for
the seed galactic dynamo field, e.g. Chakrabarti, Rosner,
Vainshtein~1994).

\begin{table}
\begin{center}
\begin{tabular}{llll|ll} 
 & $r_{\rm in} $ & L & M & $\tau_{eq}$ & $B_{eq}$ \\ \hline\hline
Black Hole & $3\cdot 10^6$ 
& $10^4$ & 10 & $7\cdot 10^{-4}$ & $5\cdot 10^{7}$\\ 
Neutron Star & $10^6$ & $10^4$ & 1 & $10^{-3}$ & $2\cdot 10^{7}$ \\
White Dwarf & $10^8$ & $10$ & 1 & $2\cdot 10^{4}$ & $2\cdot 10^{4}$ \\ 
Protosun &
$7\cdot 10^{9}$ & $332$ & $ 10^{-1}$ & $2\cdot 10^{9}$ & $8\cdot 10^{3}$\\ 
AGN & $3\cdot 10^{13}$ & $10^{12}$ &
$10^{8}$ & $7\cdot 10^{7}$ & $5\cdot 10^{3}$ \\ 
Galaxy & $10^{22}$ & $10^{10}$ &
$10^{10}$ & $6\cdot 10^{30}$ & $6\cdot 10^{-5}$ \\
\end{tabular}
\end{center}
\caption{$r_{\rm in}$ in cm; $L$ in $L_\odot$; $M$ in $M_\odot$; 
$\tau_{eq}$ in yr; $B_{eq}\equiv \Psi_{\rm eq}/r_{\rm in}^2$ in G.}
\end{table}

When we apply our mechanism to the most interesting stellar case,
namely that of the Sun, we consider a scenario in which the innermost,
optically thick, hydrogen burning core grows toward its main sequence
values $r_c\sim 0.1 r_\odot$, and $M_c\sim 0.1 M_\odot$ respectively,
or $x \simeq 2 \times 10^5$. We then consider that the remainder of
the stellar mass ($ \simeq 1$ M$_{\odot}$) accretes onto this
component in the form of an ADAF, releasing and radiating away its
gravitational energy.  In this respect, the recent discovery of X-ray
emission from objects in this class (Feigelson, private communication)
is, to say the least, encouraging.  Finally, we assume that the newly
forming core radiates at some fraction $f=0.1$ of the Eddington
luminosity that corresponds to a mass equal to $0.1M_\odot$ (or
equivalently $L=332~L_\odot$; Casanova~{\em et al.}~1995; Gross~{\em
et al.}~1998).  The above are supposed to be crude approximations
during the final stages of almost free-fall protostellar collapse.  In
the general protostellar collapse phase,
\begin{equation}
\Psi\simeq 0.4\Psi_\odot\ fM_{-1}^{1.5} (r_c/0.1\
r_\odot)^{-1.5}(t/10^6~{\rm yr})\ .
\end{equation}
Note that because of the linearity between the mass and the radius
of the initial (proto)stellar cores, the above estimate, in 
accordance with eq.(\ref{psixt}), is independent of their precise 
values, depending only on their ratio $x$. 
As one can see, for $f \sim 0.1$, it takes several million years to
build the present day total observed solar dipolar magnetic flux
$\Psi_\odot=1~{\rm G}\cdot r_\odot^2$. Considering the simplicity of
these arguments, the agreement with observation is not unreasonable.

We would like to discuss what is the geometry of the field thus
generated and its fate after the mechanism ceases to be important
(i.e. after the optically thin accretion flow
 dies out): The field is contained at the very
core, in a layer around $r_c$.  In the absence of turbulence, the
diffusion time--scale is of the order of several billion years,
and the field will remain effectively trapped inside the solar
radiation zone, i.e. below the turbulent surface convection zone.  Any
small fraction of this field which diffuses out of the sun will assume
an overall dipolar geometry at large scales. In our mechanism the
field is generated in situ, rather than being brought in from large
distances.  This magnetic field geometry should naturally interact
with stellar differential rotation, leading thus to torsional
oscillations which might account for the solar cycle (e.g. Layzer,
Rosner \& Doyle~1979; Goode \& Dziembowski~1991).

\section{Discussion}

The scenario described in this paper is not new. Previous researchers
considered the rotation of solar coronal layers (Cattani \&
Sacchi~1966) as the source of the solar magnetic field, 
and the rotation of protogalaxies with respect to the
cosmic microwave background radiation (Harrison 1970; 
Mishustin \& Ruzmaikin~1972), as a means for producing the observed
galactic field and (correctly) obtained very weak magnetic 
field values in both cases. However, to the best of our knowledge, 
this is the first time that this scenario has been applied to other 
classes of objects, in particular to optically
thin accretion flows onto compact objects, where it appears to have its
greatest impact, as the figures in table 1 indicate. 

Concerning the galactic magnetic fields, or better
the value of the galactic magnetic flux $\Psi_g \sim 3 \times 10^{36}
~ {\rm G ~cm}^2$ (this is mainly azimuthal and confined on the plane
of the galaxy) which presents the largest discord between the
predictions of our model and observations (see table 1), 
one could conceivably 
consider its generation during an early AGN phase of our galaxy, 
instead of the era of galaxy formation; as such one could argue for 
the generation of the requisite flux near a massive black hole at 
the galactic center and its diffusion to large distances
as discussed in the previous section, or as invoked by Chakrabarti, 
Rosner \& Vainshtein (1994). Assuming that such an AGN stage lasted
$\sim 10^9$ yrs, i.e. a few times the Eddington time (a quantity
which is independent of the mass of the compact object), during  
which the mass galactic center black hole presumably grew to its 
present size and the associated magnetic field 
reached its equipartition value, 
eq.(\ref{psixt}) yields a poloidal flux roughly $10^4$ times smaller
than that observed. However, the galactic differential rotation 
in the ensuing $10^{10}$ years (corresponding to roughly
100 galactic rotations at the outer regions of the galaxy and 
100 - 300 times this value for its inner regions) could easily  amplify 
this figure to its presently observed value!

Considering in particular the generation of the fields in stars, this
scenario can yield observationally interesting values for the magnetic
field only if one is willing to consider that during the 
pre--main sequence collapse of the cloud out of which the
star formed, an advection--dominated accretion flow was probably
present, and only if one is willing to accept values for the 
luminosity of this object (as given by $f$) and the duration of 
this process similar to those given above. However, even
for  lower values for $f$ or for times shorter than $\sim 10^7$ 
years, as in the above galactic case, the magnetic flux in the 
stellar interior could subsequently be amplified linearly with 
time by the interior differential rotation to observationally 
interesting values. Given the very general scope of this work 
we would not like to pursue these issues beyond this level at 
present. We feel however that the assumptions about the 
pre--main sequence protostellar evolution employed in our arguments 
are not unreasonable.  Interestingly enough, were the stellar
magnetic field to be generated as discussed in the previous 
section, it should be strong and largely confined to the deep stellar
interior, leading naturally to the torsional oscillations between
differentially rotating stellar layers which have been purported to
account for stellar cycles.

One cannot fail to notice that, when the outside magnetosphere can
support electric currents (i.e. when there is no vacuum outside), the
magnetic field geometry obtained in our numerical simulations, leads
naturally to the electromagnetic interaction between the central
region and the extended disk. The reason is that the differential
rotation `twists' the newly formed field, generating thus magnetic
torques on the central part. The magnetic field twisting leads to
another very important effect, namely the overall opening of the
closed loop geometry (Newman, Newman \& Lovelace~1992; Goodson,
Winglee \& B\"ohm~1997).\footnote{In a simple `mechanistic'
interpretation, this can be thought of as the dynamical spring--like
release through unwinding, of the azimuthal magnetic field generated
by the differential rotation of the magnetic field footpoints.}  As
one can see in the above references, the axial part of the expanding
plasma~$+$~magnetic field remains well collimated.  Under certain
physical circumstances (Shibata, Tajima \& Matsumoto~1992; Matsumoto,
Matsuzaki, Tajima \& Shibata~1997), this effect might also proceed
explosively, leading thus to the expulsion of a significant fraction
of the disk material in the form of collimated fast axially moving
plasma outflows (the astrophysical plasma gun, Contopoulos~1995;
Matsumoto {\em et al.}~1996).

In this respect it is of interest that the time necessary for the
magnetic field to become dynamically important is larger than the
Eddington time for the most massive AGN with $M\sgreat 3\cdot
10^8~M_\odot$. Thus one could speculate that the bright, optically
selected QSOs have not had yet time to generate dynamically
significant amounts of magnetic flux. One could further speculate that
when they do so, the associated magnetic field would become an
important component of the dynamics of accretion, facilitating the
production of radio jets and leading to conversion of these objects
from radio quiet to radio loud AGN.

The evolution of AGN and in particular its possible relation to their
radio properties are far from being understood. The arguments put
forward herein, associated with the generation and evolution of their
magnetic fields may offer a new way of looking into this
problem. Interestingly, the models of Falcke \& Biermann~(1995) which
provide scalings of the radio loud objects from the quasar to the
galactic binary source domain, do depend, to a large extent,  on
equipartition arguments such as those discussed above. It would be
worthwhile to examine whether these arguments could be put in the
context of an evolutionary scenario of both the mass or luminosity and
the associated magnetic field.

To conclude, it seems that the mechanism discussed above enters as a new,
previously unaccounted for, player in the ongoing studies of pre--main
sequence stellar collapse, the generation of stellar and 
galactic magnetic fields, the spin evolution of accreting magnetic 
stars, and the generation of magnetically driven jets.

\acknowledgements{We would like to thank the anonymous referee
for his constructive, penetrating criticism which, inducing a major
revision to an earlier version of this work, contributed 
significantly to its improvement. D.K. would like to acknowledge an informative discussion
with E. Feigelson. I.C. wishes to acknowledge support by grant 107526
of the General Secretariat of Research and Technology of Greece.}

\newpage
\section*{Figure Captions}

{\bf Fig. 1.}--- Magnetic field evolution across the rotating
plasma. In (a), we plot the magnetic field $b=B_z/B_o$ distribution as
a function of the dimensionless radius $r/r_{\rm in}$, at time
intervals $10 t_o$, for ${\cal P}_m=2.5$.  In (b), we plot the growth
of the magnetic flux $\psi=\Psi/(r_{\rm in}^2B_o)$ contained inside
$r=r_{\rm in}$ with time, for different Prandtl numbers ${\cal
P}_m$. One sees clearly the fast growth of the field around $r_{\rm
in}$, and the almost linear growth of the central magnetic flux for
${\cal P}_m>2$.  The present (kinematic) integration should not be
continued beyond $\tau_{eq}$, when the field becomes dynamically
important.

\begin{figure}
\plotone{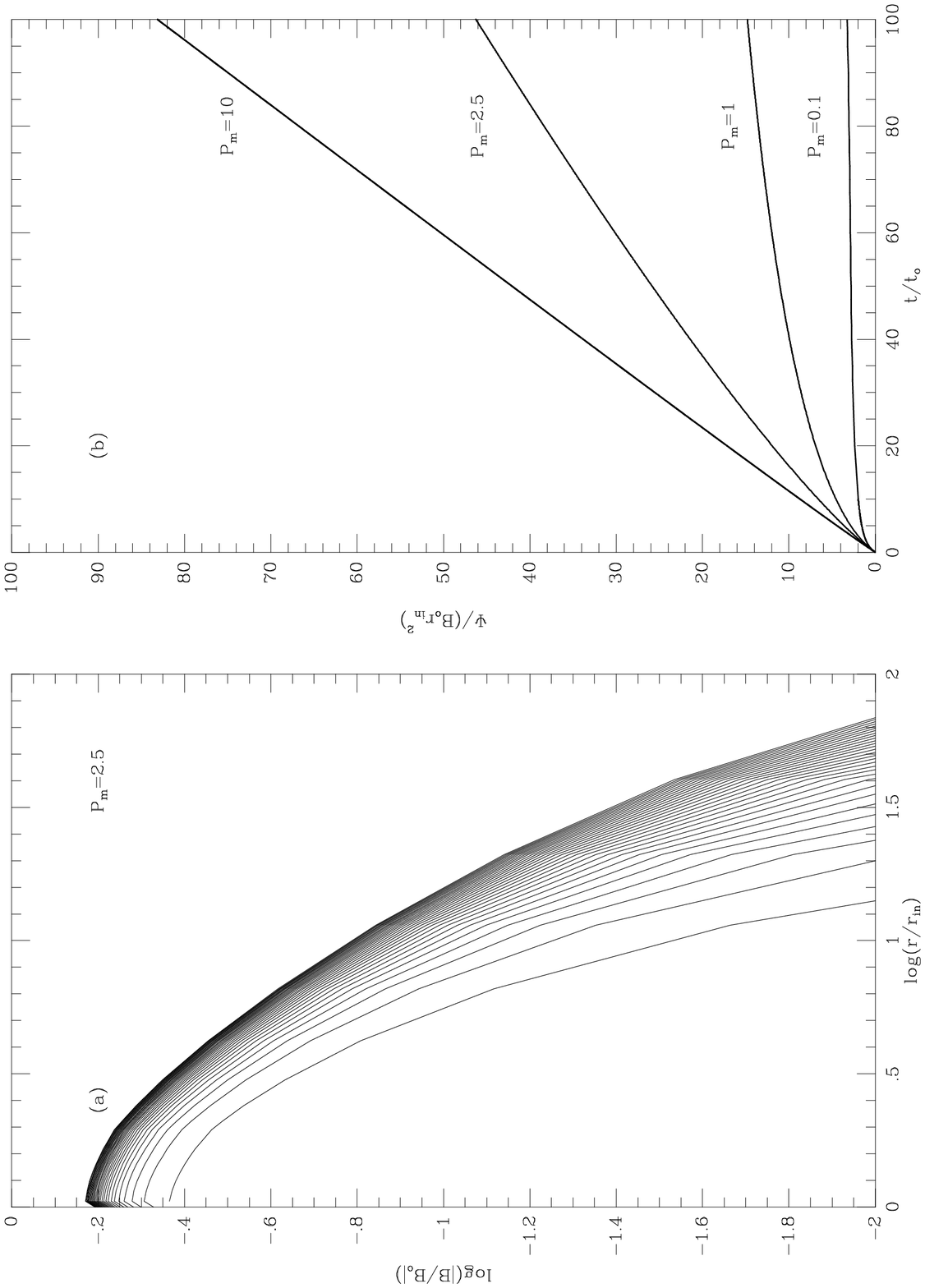}
\caption{}
\end{figure}

\end{document}